\documentclass[12pt]{article}

\usepackage{latexsym}
\usepackage{amssymb,amsfonts,amsmath}
\usepackage{graphicx} 
\usepackage{indentfirst}
\usepackage{bbm}
\usepackage{amssymb}
\usepackage{verbatim}
\usepackage{amsmath, amsthm,amssymb}
\usepackage{mathrsfs}
\usepackage{hyperref}
\usepackage{amsfonts}
\usepackage{dsfont}
\usepackage{cite}
\usepackage{xcolor}
\usepackage[multiple]{footmisc}

\parskip=\medskipamount

\newcommand {\cD}{{\cal D}}
\newcommand {\cE}{{\cal E}}

\newcommand {\cM}{{\cal M}}
\newcommand {\cN}{{\cal N}}

\def\a{\alpha}

\def\b{\beta}

\def\d{\delta}

\def\g{\gamma}
\def\G{\Gamma}

\def\l{\lambda}
\def\m{\mu}

\def\o{\omega}

\def\q{\theta}
\def\r{\rho}

\def\z{\zeta}
\def\D{\Delta}

\def\J{\Psi}
\def\L{\Lambda}

\def\U{\Upsilon}

\def\rd{{\rm d}}
\def\ri{{\rm i}}
\def\re{{\rm e}}

\newcommand{\ad}{{\dot{\alpha}}}                           
\newcommand{\bd}{{\dot{\beta}}}                            
\newcommand{\ve}{\varepsilon}                            

\newcommand{\ab}{{\a\b}}

\renewcommand{\aa}{{\a\ad}}
\newcommand{\bb}{{\b\bd}}
\newcommand{\pa}{\partial}                           
\newcommand{\hf}{\frac12}

\newcommand{\be}{\begin{equation}}
\newcommand{\ee}{\end{equation}}
\newcommand{\bea}{\begin{eqnarray}}
\newcommand{\eea}{\end{eqnarray}}
\newcommand{\non}{\nonumber}
\newcommand{\bm}[1]{\mbox{\boldmath$#1$}}

\def\double #1{#1{\hbox{\kern-2pt $#1$}}}

\newcommand{\gd}{{\dot\g}}
\newcommand{\dd}{{\dot\d}}

\newcommand{\Nabla}{\bm{\nabla}}
\newcommand{\bNabla}{\bar{\bm{\nabla}}}

\newcommand{\dalpha}{{\dot{\alpha}}}

\newcommand{\N}{{\mathcal N}}
\newif\ifdtup

\newcommand{\bsubeq}{\begin{subequations}}
\newcommand{\esubeq}{\end{subequations}}

\newcommand{\eol}{\notag \\}

\numberwithin{equation}{section}

\newcommand{\sSU}{\mathsf{SU}}

\newcommand{\sU}{\mathsf{U}}

\newcommand{\sOSp}{\mathsf{OSp}}

\begin{document}

\begin{titlepage}
\begin{flushright}
April, 2021 \\
Revised version: December, 2021\\
\end{flushright}
\vspace{5mm}

\begin{center}
{\Large \bf Extended superconformal higher-spin gauge theories in four dimensions}
\end{center}

\begin{center}

{\bf Sergei M. Kuzenko and Emmanouil S. N. Raptakis} \\
\vspace{5mm}

\footnotesize{
{\it Department of Physics M013, The University of Western Australia\\
35 Stirling Highway, Perth W.A. 6009, Australia}}  
~\\
\vspace{2mm}
~\\
Email: \texttt{ 
sergei.kuzenko@uwa.edu.au, emmanouil.raptakis@research.uwa.edu.au}\\
\vspace{2mm}

\end{center}

\begin{abstract}
\baselineskip=14pt
Using the off-shell formulation for ${\cal N}=2$ conformal supergravity in four dimensions, we describe superconformal higher-spin multiplets of conserved currents in a curved background and present their associated unconstrained gauge prepotentials. The latter are used to construct locally superconformal chiral actions, which are demonstrated to be gauge invariant in arbitrary conformally flat backgrounds. The main $\cN=2$ results are then generalised to the $\cN$-extended case. We also present the gauge-invariant field strengths for on-shell massless  higher-spin $\cN=2$ supermultiplets in anti-de Sitter space. These field strengths prove to furnish representations of the $\cN=2$ superconformal group.
\end{abstract}
\vspace{5mm}

\vfill

\vfill
\end{titlepage}

\newpage
\renewcommand{\thefootnote}{\arabic{footnote}}
\setcounter{footnote}{0}

\tableofcontents{}
\vspace{1cm}
\bigskip\hrule

\allowdisplaybreaks


\section{Introduction}

In four dimensions, 
off-shell $\cN=1$ superconformal higher-spin prepotentials
were briefly discussed by Howe, Stelle and Townsend 
 in the framework of supercurrent multiplets in 1981 \cite{HST},  
 a few years before Fradkin and Tseytlin \cite{FT} 
constructed the free conformal higher-spin actions in Minkowski space.
It was only in the last few years that 
these prepotentials and more general off-shell gauge supermultiplets 
were finally used to construct free $\cN=1$ superconformal higher-spin gauge theories
in Minkowski space \cite{KMT}
and arbitrary conformally flat and Bach flat backgrounds \cite{KP,KR19,KPR1,KPR2}.
Parallel studies in three dimensions (3D) to describe superconformal higher-spin
 multiplets and the corresponding Chern-Simons actions 
 were conducted \cite{KO,K16,KTsulaia,SK-MP,HKO,BHHK}.
These 3D and 4D off-shell constructions 
open the possibility to develop a manifestly supersymmetric setting for 
superconformal higher-spin gauge theories which were considered for the first time  
by Fradkin and Linetsky \cite{FL-3D,FL-4D} in the component approach.\footnote{The Fradkin-Linetsky formulation  \cite{FL-3D,FL-4D} may be viewed as a natural extension of the Fradkin-Vasiliev approach to massless higher-spin gauge theories \cite{FV,Vasiliev}.}
 
 One of the goals of this paper is to extend some of the results of \cite{KMT,KP,KR19,KPR1} to the 4D $\cN=2$ superconformal case.
 Using the $\cN=2$ conformal superspace approach \cite{ButterN=2}, we will propose superconformal higher-spin multiplets of conserved currents and their associated unconstrained gauge prepotentials. The corresponding superconformal gauge-invariant actions will be constructed in arbitrary conformally flat backgrounds.
 We will also generalise the main $\cN=2$ results to the $\cN$-extended case. 
 
 Our approach to determine the structure of superconformal higher-spin gauge prepotentials is to use the method of supercurrent multiplets \cite{HST,OS,BdRdW}.
 It is well known that the multiplets of (conformal) currents furnish off-shell representations of (conformal) supersymmetry. Once a conformal higher-spin supercurrent $J$ is known,
 the associated gauge prepotential $\U$ is determined via the Noether coupling
\bea
\label{NC}
S_{\text{NC}} = \int \rd^4x \rd^4 \q \rd^4 \bar\q \, E\, \U \cdot J ~.
\eea
This procedure is concisely described
by Bergshoeff et {\it al.} \cite{BdRdW}:
``One first assigns a field to each component of the current multiplet,
and forms a generalized inner product of field and current components.''

In this work we make use of $\cN=2$ conformal superspace \cite{ButterN=2}, which is an ultimate formulation for $\cN=2$ 
conformal supergravity in the sense that any different off-shell formulation is either equivalent to it or is obtained from it by partially fixing the gauge freedom. 
In particular,  $\sU(2)$ superspace \cite{Howe2} and $\sSU(2)$ superspace\cite{KLRT-M1} can be derived from conformal superspace by 
imposing partial gauge fixing conditions.\footnote{The relationship between  the  $\sU(2)$ and $\sSU(2)$ superspaces is described 
	in \cite{KLRT-M2}.}
At the component level, $\cN=2$ conformal superspace reduces to 
the $\cN=2$ superconformal tensor calculus \cite{BdRdW,deWvHVP1,deRvHdeWVP}.
The recently discovered supertwistor formulation for 
$\cN=2$ supergravity \cite{HL20} is believed to be equivalent to conformal superspace, however the technical details are yet to be worked out.

This paper is organised as follows. In section \ref{section2} we determine all $\N = 2$ conformal supercurrents. This in turn allows us to compute their dual gauge prepotentials and corresponding gauge invariant actions in conformally flat backgrounds in section \ref{section3}. Our key $\cN=2$ results are generalised 
to the $\cN$-extended case in section \ref{section4}.
Concluding comments and implications of the obtained results are given in section \ref{section5}. The main body of this paper is accompanied by a technical appendix 
reviewing the relevant aspects of $\cN=2$ conformal superspace.


\section{Conformal higher-spin supercurrents} \label{section2}

The objective of this section is to identify all possible $\N=2$ conformal higher-spin supercurrents $J$ in a curved background, which will in turn elucidate the structure of their dual gauge prepotentials $\U$. The technical details regarding $\cN=2$ conformal superspace are given in the appendix. 

\subsection{$\mathcal{N}=2$  conserved current supermultiplets}

Let $m$ and $n$ be positive integers. A primary tensor superfield $J^{\a(m) \ad(n)}$ defined on the background superspace will be called a conformal supercurrent if it obeys
\begin{subequations}
\label{SC1}
\bea
\nabla_\b^i J^{\b \a(m-1) \ad(n)} &=& 0 \quad \Longrightarrow \quad \nabla^{ij} J^{\a(m) \ad(n)} = 0 ~, \\
\bar{\nabla}_\bd^i J^{\a(m) \bd \ad(n-1)} &=& 0 \quad \Longrightarrow \quad \bar{\nabla}^{ij} J^{\a(m) \ad(n)} = 0 ~,
\eea
\end{subequations}
where we have denoted
\bea
{\nabla}^{ij} = {\nabla}^{\a(i} {\nabla}_\a^{ j)} ~, \qquad
\bar{\nabla}^{ij} = \bar{\nabla}_{\ad}^{(i} \bar{\nabla}^{\ad j)} 
 ~.
\eea
These constraints uniquely fix the superconformal properties of $J^{\a(m) \ad(n)}$
\bea
\mathbb{D} J^{\a(m) \ad(n)} = \hf (m+n+4) J^{\a(m) \ad(n)} ~, \quad Y J^{\a(m) \ad(n)} = -(m-n) J^{\a(m) \ad(n)} ~.
\eea
For $m=n=s$, $J^{\a(s) \ad(s)}$ is invariant under $\sU(1)_R$ transformations and thus we take it to be real. This special case was first described in Minkowski superspace in \cite{HST}.

When $n = 0$, the constraints \eqref{SC1} should be replaced with
\begin{subequations}
\label{SC2}
\bea
\nabla_\b^i J^{\b \a(m-1)} &=& 0 \quad \Longrightarrow \quad \nabla^{ij} J^{\a(m) } = 0 ~, \\
\bar{\nabla}^{ij} J^{\a(m)} &=& 0 ~.
\eea
\end{subequations}
Consistency of \eqref{SC2} with the superconformal algebra implies:
\bea
\mathbb{D} J^{\a(m)} = \hf (m + 4) J^{\a(m)} ~, \quad Y J^{\a(m)} = -m J^{\a(m)} ~.
\eea

Finally, for the special case $m=0$ , we replace \eqref{SC2} with
\label{SC3}
\bea
\nabla^{ij} J = 0 ~, \quad \bar{\nabla}^{ij} J = 0 ~.
\eea
These imply:
\bea
\mathbb{D} J = 2 J ~, \quad Y J= 0 ~.
\eea
Taking $J$ to be real, it is clear that it corresponds to the conformal supercurrent
\cite{HST,Sohnius,KT}.

Analogous to the $\N=1$ case \cite{KR19}, conformal Killing tensors may be utilised to construct new conserved conformal  currents from existing ones. The former are primary tensor superfields $\z_{\a(p) \ad(q)}$, $p,q \geq 0$, satisfying
\bea
\nabla_{(\a_1}^i \z_{\a_2 \dots \a_{p+1}) \ad(q)} = 0 ~, \quad \bar \nabla_{(\ad_1}^i \z_{\a(p) \ad_2 \dots \ad_{q+1})} = 0 ~.
\eea
In particular, given a conformal supercurrent $J^{\a(m)\ad(n)}$ and a conformal Killing tensor $\z_{\a(p) \ad(q)}$, with $m \geq p$ and $n \geq q$, it may be shown that
\bea
\mathfrak{J}^{\a(m-p) \ad(n-q)} = J^{\a(m-p) \b(p) \ad(n-q) \bd(q)} \z_{\b(p) \bd(q)} ~,
\eea
is also a conformal supercurrent. The special cases $p=q$ and $p=q=2$ were introduced in \cite{HL1} and \cite{HL3}, respectively. In a forthcoming work, we will further explore their properties, in particular, their role in the study of higher symmetries \cite{KRWIP}, see also \cite{HL2}.


\subsection{Reduction to $\N=1$ superspace}

To conclude our discussion of  the conformal higher-spin supercurrents,  it is instructive to comment on the 
$\N=1$ supermultiplets contained within the conformal supercurrents described above. This analysis requires us to turn off the super-Weyl tensor, that is:
\bea
\label{cflat}
W_{\ab} = 0.
\eea
Let $\Nabla_\a $, $\bar{\Nabla}^\ad $ and $\Nabla_{\aa} = \frac{\ri}{2} \{\Nabla_\a , \bar \Nabla_\ad \}$ be the covariant derivatives of $\N=1$ conformal superspace \cite{ButterN=1}. We will define them using the $\cN=2$ covariant derivatives
 \eqref{A.1} as follows: $ \Nabla_\a {\mathfrak U} =  \nabla_\a^{\underline{1}} U|$
and $\bar{\Nabla}^\ad {\mathfrak U}= \bar{\nabla}^{\ad}_{\underline{1}} U|$.
Here $U$ is an $\cN=2$ superfield, and  ${\mathfrak U} \equiv U| :=U|_{\theta^\a_{\underline 2} = \bar{\theta}_\ad^{\underline 2} = 0}$ is its $\N=1$ projection. We will say that $\mathfrak U$ is primary if $\bold{S}^\a \mathfrak U \equiv S^{\a}_{\underline{1}} U | = 0$ and $\bar{\bold{S}}_\ad \mathfrak U \equiv \bar{S}_{\ad}^{\underline{1}} U | = 0$. Using these definitions, one may derive the important identities
\bea
\{ \bold{S}^\a , \Nabla_\b \} = \d^{\a}_\b (2 \mathbb{D} - 3 \mathbb{Y}) - 4 M^{\a}{}_{\b} ~, \quad \{ \bar{\bold{S}}_\ad , \bNabla^\bd \} = \d_{\ad}^\bd (2 \mathbb{D} + 3 \mathbb{Y}) + 4 \bar M_{\ad}{}^{\bd}~,
\eea
where $\mathbb{Y}$ is the $\sU(1)_R$ generator. It is related to the $\mathcal{N}=2$ R-symmetry generators via
\bea
\mathbb{Y} = \frac{1}{3} Y - \frac{4}{3} J_{\underline{1}}{}^{\underline{1}} ~.
\eea

We now review the key properties of $\N = 1$ conformal current supermultiplets (see \cite{Ceresole:1999zs} and \cite{KR19} for a complete discussion in flat and curved backgrounds, respectively). A primary tensor superfield $\mathcal J^{\a(m) \ad(n)}$, with $m,n \geq 1$, obeying the constraints
\begin{subequations}\label{supercurrent} 
\bea
\Nabla_{\b} \mathcal J^{\b\a (m-1)\ad(n)} &=& 0 \quad \implies \quad \Nabla^2 \mathcal J^{\a(m) \ad(n)} = 0 ~, 
\label{supercurrent-a} 
\\
\bar \Nabla_{\bd} \mathcal J^{\a(m) \bd \ad(n-1) } &=& 0 \quad \implies \quad \bar \Nabla^2 \mathcal J^{\a(m) \ad(n)} = 0 ~,
\label{supercurrent-b} 
\eea
\end{subequations}
is a conformal supercurrent. 
 The $m=n=1$ case corresponds to the ordinary 
conformal supercurrent \cite{FZ}. 
For case $m >n =0$, the appropriate constraints are 
\begin{subequations}\label{supercurrent2} 
\bea
\Nabla_{\b} \mathcal J^{\b\a (m-1)} &=& 0 \quad \implies \quad \Nabla^2 \mathcal J^{\a(m)} = 0 ~, 
\label{supercurrent2-a} 
\\
\bar \Nabla^2 \mathcal J^{\a(m)  } &=& 0~.
\label{supercurrent2-b} 
\eea
\end{subequations}
The  $m=1$ case was first considered in \cite{KT}, where it was shown that the spinor supercurrent $J^\a$ naturally originates from the reduction of the conformal $\cN=2$ supercurrent \cite{Sohnius}
to $\cN=1$ superspace. 
Finally, when $m=0$ the supercurrent satisfies
\bea
\label{supercurrent3} 
\Nabla^2 \mathcal J =0~, \quad 
\bar \Nabla^2 \mathcal J = 0~.
\eea
This is the flavour current supermultiplet \cite{FWZ}.

Returning to the reduction procedure, we first consider the $\N=2$ supercurrent $J^{\a(m)\ad(n)}$ \eqref{SC1}. It contains four independent $\N=1$ conserved current supermultiplets
\begin{subequations}
\label{2.16}
\bea
j^{\a(m) \ad(n)} &=& J^{\a(m) \ad(n)} | ~, \\
j^{\a(m+1) \ad(n)} &=& \nabla^{(\a_1 \underline 2} J^{\a_2 \dots \a_{m+1}) \ad(n)} | ~, \\
j^{\a(m) \ad(n+1)} &=& \bar \nabla^{(\ad_1}_{\underline 2} J^{\a(m) \ad_2 \dots \ad_{n+1})} | ~, \\
j^{\a(m+1) \ad(n+1)} &=& \hf \big[ \nabla^{ (\a_1 \underline 2} , \bar \nabla_{\underline{2}}^{(\ad_1} \big] J^{\a_2 \dots \a_{m+1}) \ad_2 \dots \ad_{n+1})} | \non \\
&-& \frac{1}{2(m+n+3)} \big[ \Nabla^{(\a_1} , \bar \Nabla^{(\ad_1} \big] j^{\a_2 \dots \a_{m+1}) \ad_2 \dots \ad_{n+1})} \non \\
&-& \frac{\ri (m-n)}{m+n+3} \Nabla^{(\a_1 (\ad_1} j^{\a_2 \dots \a_{m+1}) \ad_2 \dots \ad_{n+1})} ~.
\eea
\end{subequations}
Similarly, $J^{\a(m)}$ \eqref{SC2} is composed of four $\N=1$ supercurrents
\begin{subequations}
\bea
j^{\a(m)} &=& J^{\a(m)} | ~, \\
j^{\a(m+1)} &=& \nabla^{(\a_1 \underline 2} J^{\a_2 \dots \a_{m+1})} | ~, \\
j^{\a(m) \ad} &=& \bar \nabla^{\ad}_{\underline 2} J^{\a(m)} | ~, \\
j^{\a(m+1) \ad} &=& \hf \big[ \nabla^{(\a_1 \underline 2} , \bar \nabla_{\underline{2}}^{\ad} \big] J^{\a_2 \dots \a_{m+1})} | - \frac{1}{2(m+3)} \big[ \Nabla^{(\a_1} , \bar \Nabla^{\ad} \big] j^{\a_2 \dots \a_{m+1})} \non \\
&-& \frac{\ri m}{m+3} \Nabla^{(\a_1 \ad} j^{\a_2 \dots \a_{m+1})} ~.
\eea
\end{subequations}
Finally, upon reduction of $J$ \eqref{SC3} we obtain three $\N=1$ current multiplets \cite{KT}
\begin{subequations}
\bea
j^{} &=& J^{} | ~, \\
j^{\a} &=& \nabla^{\a \underline 2} J | ~, \\
j^{\a \ad} &=& \hf \big[ \nabla^{\a \underline 2} , \bar \nabla_{\underline{2}}^{\ad} \big] J | - \frac{1}{6} \big[ \Nabla^{\a} , \bar \Nabla^{\ad} \big] j ~.
\eea
\end{subequations}
For further details regarding their dual gauge prepotentials and the superconformal field theories they induce, we refer the reader to \cite{KR19,KP,KMT,KPR1}. See also \cite{KPR2} for their higher-depth generalisations.


\section{$\cN=2$ superconformal higher-spin gauge models} \label{section3}

In the previous section, we described the $\N=2$ conformal higher-spin supercurrents. Here, by requiring that the Noether coupling \eqref{NC} is superconformal and gauge-invariant, we will identify their dual gauge prepotentials.

The first family of supercurrents take the form $J^{\a(m) \ad(n)}$, $m,n \geq 1$, and are subject to \eqref{SC1}. It then follows that their duals, $\U_{\a(m) \a(n)}$, are defined modulo gauge transformations
\bea
\label{SCHSgt1}
\d_{\z,\l} \U_{\a(m) \ad(n)} = \nabla_{(\a_1}^i \z_{\a_2 \dots \a_m) \ad(n) i} + \bar{\nabla}_{(\ad_1}^i \l_{\a(m) \ad_2 \dots \ad_n) i} ~,
\eea
where $\z_{\a(m-1) \ad(n) i}$ and $\l_{\a(m) \ad(n-1) i }$ are complex unconstrained.

When $n=0$, we obtain the second family of supercurrents, namely $J^{\a(m)}$, which satisfy \eqref{SC2}. It is clear that the corresponding prepotentials, $\U_{\a(m)}$, are characterised by the gauge transformation law:
\bea
\label{SCHSgt2}
\d_{\z,\o} \U_{\a(m)} = \nabla_{(\a_1}^i \z_{\a_2 \dots \a_m) i} + \bar{\nabla}^{ij} \o_{\a(m) ij} ~,
\eea
where $\z_{\a(m-1) i}$ and $\o_{\a(m) ij }$ are complex unconstrained.

Finally, for $m=0$, we obtain the real scalar supercurrent $J$. It is constrained by \eqref{SC3} and its dual, $\U$, may be chosen to be real and has the gauge transformation law
\bea
\label{SCHSgt3}
\d_{\o} \U = \bar \nabla^{ij} \o_{ij} + \text{c.c.}~,
\eea
where the gauge parameter $\o_{ij}$ is complex unconstrained. As expected, this prepotential describes the conformal supergravity multiplet.

\subsection{Superconformal models for the $\U_{\a(m) \ad(n)}$ prepotentials}

To begin with, we study the prepotentials $\U_{\a(m) \ad(n)}$ \eqref{SCHSgt1}. Requiring that both these and their corresponding gauge parameters, $\z_{\a(m-1) \ad(n) i}$ and $\l_{\a(m) \ad(n-1) i }$, are superconformally primary,
\bea
K^B \U_{\a(m) \ad(n)} = 0 ~, \quad K^B \z_{\a(m-1) \ad(n) i} = 0 ~, \quad K^B \l_{\a(n) \ad(n-1) i } = 0 ~,
\eea
uniquely fixes the dimension and $\sU(1)_R$ charge of $\U_{\a(m) \ad(n)}$
\bea
\mathbb{D} \U_{\a(m) \ad(n)} = - \hf (m+n+4) \U_{\a(m) \ad(n)} ~, \quad Y \U_{\a(m) \ad(n)} = (m - n) \U_{\a(m) \ad(n)} ~.
\eea
This implies that for $m = n = s$, we are able to choose $\U_{\a(s) \ad(s)}$ to be real, in which case \eqref{SCHSgt1} reduces to
\bea
\label{realPPgt}
\d_\z \U_{\a(s) \ad(s)} = \nabla_{(\a_1}^i \z_{\a_2 \dots \a_s) \ad(s) i} - \bar{\nabla}_{(\ad_1}^i \bar{\z}_{\a(s) \ad_2 \dots \ad_s) i} ~.
\eea
In flat superspace, \eqref{realPPgt} was first proposed in \cite{HST}.

From the prepotential $\U_{\a(m) \ad(n)}$ and its conjugate $\bar{\U}_{\a(n) \ad(m)}$, we may construct the higher-derivative chiral descendants\footnote{In the flat-superspace limit, these chiral field strengths reduce to those introduced in \cite{SG,GGRS}. }
\begin{subequations}
\label{linSW1}
\bea
\hat{\mathfrak{W}}_{\a(m+n+2)} (\U) &=& \bar \nabla^4 \nabla_{(\a_1}{}^{\bd_1} \dots \nabla_{\a_n}{}^{\bd_n} \nabla_{\a_{n+1} \a_{n+2}} \U_{\a_{n+3} \dots \a_{m+n+2}) \bd(n)} ~, \\
\check{\mathfrak{W}}_{\a(m+n+2)} (\bar \U) &=& \bar \nabla^4 \nabla_{(\a_1}{}^{\bd_1} \dots \nabla_{\a_m}{}^{\bd_m} \nabla_{\a_{m+1} \a_{m+2}} \bar \U_{\a_{m+3} \dots \a_{m+n+2}) \bd(m)} ~.
\eea
\end{subequations}
Here we have introduced 
the chiral projection operator 
\bea
\bar{\nabla}^4  \equiv \frac{1}{48} \bar{\nabla}^{ij} \bar{\nabla}_{ij} ~
\eea
and the second-order operators 
\begin{align}
\nabla_{\a\b} := \nabla_{(\a}^k \nabla_{\b) k} \ , \qquad
 \bar{\nabla}^{\ad\bd} := \bar\nabla^{(\ad}_k \bar\nabla^{\bd) k} \ .
\end{align}
It should be emphasised that, for the special case $m=n=s$, the chiral descendants \eqref{linSW1} coincide; $\hat{\mathfrak{W}}_{\a(2s+2)} (\U) = \check{\mathfrak{W}}_{\a(2s+2)} (\bar{\U}) \equiv \mathfrak{W}_{\a(2s+2)} (\U)$.
The chiral field strengths have the following dimensions:
\begin{subequations} \label{3.100}
\bea
\mathbb{D} \hat{\mathfrak{W}}_{\a(m+n+2)} (\U) &=&  \hf (n-m +2) \hat{\mathfrak{W}}_{\a(m+n+2)} (\U) ~, \\
\mathbb{D} \check{\mathfrak{W}}_{\a(m+n+2)} (\bar \U) &=& \hf (m-n + 2) \check{\mathfrak{W}}_{\a(m+n+2)}  (\bar \U) ~.
\eea
\end{subequations}
Further, it may be shown that they are primary,
\bea
K^B \hat{\mathfrak{W}}_{\a(m+n+2)} (\U) = 0 ~, \quad K^B \check{\mathfrak{W}}_{\a(m+n+2)} (\bar \U) = 0 ~.
\eea

These properties imply that the action
\bea
\label{SCHSaction1}
S^{(m,n)} = \ri^{m+n} \int \rd^4x \rd^4 \q \, \cE\, \hat{\mathfrak{W}}^{\a(m+n+2)} (\U) \check{\mathfrak{W}}_{\a(m+n+2)} (\bar \U) + \text{c.c.}
\eea
is locally superconformal. Here $\cE$ is the chiral integration measure. 
Chiral and full superspace integrals are related according to the rule
\bea
 \int \rd^4x\rd^4\theta\rd^4\bar\theta\, E\,  U
 =\int \rd^4x \rd^4\theta\, \cE\, \bar \nabla^4 U~,\qquad E^{-1} = {\rm Ber}(E_A{}^M)~,
\eea
where $U$ is a real primary dimension-0 superfield.
Derivations of this result using superspace normal coordinates are given in 
\cite{ButterN=2,KTM08}.
The overall factor of $\ri^{m+n}$ in \eqref{SCHSaction1} has been chosen due to the identity
\bea
\ri^{m+n+1} \int \rd^4x \rd^4 \q \, \cE\, \hat{\mathfrak{W}}^{\a(m+n+2)} (\U) \check{\mathfrak{W}}_{\a(m+n+2)} (\bar \U) + \text{c.c.} = 0~,
\eea
which holds up to a total derivative for any conformally flat background \eqref{cflat}. 

We restrict our attention to conformally flat superspaces. In these geometries it may be shown that the chiral descendants \eqref{linSW1} are gauge-invariant
\bea
\d_{\z,\l} \hat{\mathfrak{W}}_{\a(m+n+2)} (\U) = 0 ~, \quad \d_{\z,\l} \check{\mathfrak{W}}_{\a(m+n+2)} (\bar \U) = 0 ~.
\eea
As a result, the action \eqref{SCHSaction1} is gauge-invariant, thus the field strengths \eqref{linSW1} are linearised higher-spin super-Weyl tensors.


\subsection{Superconformal models for the $\U_{\a(m)}$ prepotentials}

The gauge prepotentials $\U_{\a(m)}$, with $m \geq 1$ are defined modulo the transformations \eqref{SCHSgt2}. Similar to the $n=0$ case, requiring that both the prepotentials and gauge parameters are superconformally primary,
\bea
K^B \U_{\a(m)} = 0 ~, \quad K^B \z_{\a(m-1) i} = 0 ~, \quad K^B \o_{\a(n) ij } = 0 ~,
\eea
determines the dimension and $\sU(1)_R$ charge of $\U_{\a(m)}$
\bea
\mathbb{D} \U_{\a(m)} = -\hf (m + 4) \U_{\a(m)} ~, \quad Y \U_{\a(m) } = m \U_{\a(m)} ~.
\eea

Associated with the prepotential $\U_{\a(m)}$ (and its conjugate $\bar{\U}_{\ad(m)}$) are the chiral descendants:
\begin{subequations}
	\label{linSW2}
	\bea
	\hat{\mathfrak{W}}_{\a(m+2)} (\U) &=& \bar \nabla^4 \nabla_{(\a_{1} \a_{2}} \U_{\a_{3} \dots \a_{m+2})} ~, \\
	\check{\mathfrak{W}}_{\a(m+2)} (\bar \U) &=& \bar \nabla^4 \nabla_{(\a_1}{}^{\bd_1} \dots \nabla_{\a_m}{}^{\bd_m} \nabla_{\a_{m+1} \a_{m+2})} \bar \U_{\bd(m)} ~.
	\eea
\end{subequations}
It may readily be shown that (i) their dimensions are
\begin{subequations}
	\bea
	\mathbb{D} \hat{\mathfrak{W}}_{\a(m+2)} (\U) &=& - \hf (m - 2)  \hat{\mathfrak{W}}_{\a(m+2)} (\U) ~,	\\
	\mathbb{D} \check{\mathfrak{W}}_{\a(m+2)} (\bar \U) &=& \hf(m + 2) \check{\mathfrak{W}}_{\a(m+2)} (\bar \U) ~;
	\eea
\end{subequations}
and that (ii) they are primary
\bea
K^B \hat{\mathfrak{W}}_{\a(m+2)} (\U) = 0 ~, \quad K^B \check{\mathfrak{W}}_{\a(m+2)} (\bar \U) = 0 ~.
\eea
As a result, the following action
\bea
\label{SCHSaction2}
S^{(m)} = \ri^{m} \int \rd^4x \rd^4 \q \, \cE\, \hat{\mathfrak{W}}^{\a(m+2)} (\U) \check{\mathfrak{W}}_{\a(m+2)} (\bar \U) + \text{c.c.}
\eea
is locally superconformal. Similar to the $n \neq 0$ case, the overall coefficient of $\ri^{m}$ has been chosen since on conformally flat backgrounds
\bea
\ri^{m+1} \int \rd^4x \rd^4 \q \, \cE\, \hat{\mathfrak{W}}^{\a(m+2)} (\U) \check{\mathfrak{W}}_{\a(m+2)} (\bar \U) + \text{c.c.} = 0~,
\eea
i.e. it is a total derivative.

In backgrounds with vanishing super-Weyl tensor, a routine calculation allows us to show that \eqref{linSW2} are gauge-invariant field strengths
\bea
\d_{\z,\l} \hat{\mathfrak{W}}_{\a(m+2)} (\U) = 0 ~, \quad \d_{\z,\l} \check{\mathfrak{W}}_{\a(m+2)} (\bar \U) = 0 ~.
\eea
Thus, just as for the $n \neq 0$ case, the action \eqref{SCHSaction2} proves to be gauge-invariant and the field strengths \eqref{linSW2} are linearised higher-spin super-Weyl tensors.


\subsection{Linearised $\N=2$ conformal supergravity}

The scalar gauge prepotential $\U = \bar \U $ describes linearised conformal supergravity  and possesses the gauge freedom \eqref{SCHSgt3}. The requirement that both $\U$ and $\o_{ij}$ are superconformal primary,
\bea
K^B \U = 0 ~, \quad K^B \o_{ ij } = 0 ~,
\eea
leads to
\bea
\mathbb{D} \U = - 2 \U ~, \quad Y \U = 0 ~.
\eea

From the prepotential $\U$, we may construct  the single chiral descendant
\bea
\label{linSW3}
\mathfrak{W}_{\a(2)} (\U) &=& \bar \nabla^4 \nabla_{\a(2)} \U ~.
\eea
It has the following superconformal properties
\bea
\mathbb{D} \mathfrak{W}_{\a(2)} (\U) =  \mathfrak{W}_{\a(2)} (\U) ~, \quad K^B \mathfrak{W}_{\a(2)} (\U) = 0 ~,
\eea
which imply that the following action 
\bea
\label{SCHSaction3}
S = \int \rd^4x \rd^4 \q \, \cE\, \mathfrak{W}^{\a(2)} (\U) \mathfrak{W}_{\a(2)} ( \U) + \text{c.c.}
\eea
is locally superconformal.
We note that 
\bea
\ri \int \rd^4x \rd^4 \q \, \cE\, \mathfrak{W}^{\a(2)} (\U) \mathfrak{W}_{\a(2)} ( \U) + \text{c.c.} = 0 ~,
\eea
is a total derivative when $W_{\a \b}=0$.

When the background geometry is conformally flat, it is easily shown that
$\mathfrak{W}_{\a(2)} (\U) $ is invariant under the gauge transformations \eqref{SCHSgt3},
\bea
\d_{\o} \mathfrak{W}_{\a(2)} (\U) = 0 ~.
\eea
Hence, \eqref{SCHSaction3} is gauge-invariant. Consequently, \eqref{linSW3} is the linearised super-Weyl tensor.

\subsection{Reduction to $\N=1$ superspace}
\label{section3.4}

This subsection is devoted to a discussion of the $\N=1$ superfield content of the
real superconformal prepotentials $\U_{\a(s) \ad(s)}$ for $s \geq 0$ in conformally-flat backgrounds.\footnote{It should be noted that for the $s=0$ case, this reduction was carried out in \cite{BKSC}.} We remind the reader that these are defined modulo the gauge transformations \eqref{realPPgt} for $s>0$ and \eqref{SCHSgt3} for $s=0$.

Utilising this freedom, we may construct a gauge on $\U_{\a(s) \ad(s)}$ such that the only non-vanishing $\N=1$ superfields in its multiplet are:
\begin{subequations}
	\label{4.16}
	\begin{align}
		\bm{H}_{\a(s+1) \ad(s+1)} &= \frac{1}{2} [\nabla_{(\a_1}^{\underline{2}} , \bar{\nabla}_{(\ad_1 \underline{2}}] \U_{\a_2 \dots \a_{s+1}) \ad_{2} \dots \ad_{s+1})} | ~, \\
		\bm{\Psi}_{\a(s+1) \ad(s)} &= - \frac{1}{4} \nabla_{(\a_1}^{\underline{2}} 	(\bar{\nabla}_{\underline{2}})^2 \U_{\a_2 \dots \a_{s+1}) \ad(s)} | ~, \\
		\bm{G}_{\a(s) \ad(s)} &= 
		\frac{1}{32}  \{ (\nabla^{\underline{2}})^2 , (\bar{\nabla}_{\underline{2}})^2 \}  \U_{\a(s) \ad(s)} | + \frac{1}{8(2s+3)} [\Nabla^{\b} , \Nabla^{\bd}] \bm{H}_{\a(s) \b \ad(s) \bd}~.
	\end{align}
\end{subequations}
We emphasise that these superfields have been defined such that they are primary
\bea
K^B \bm{H}_{\a(s+1) \ad(s+1)} = 0 ~, \qquad K^B \bm{\Psi}_{\a(s+1) \ad(s)} = 0 ~, \qquad K^B \bm{G}_{\a(s) \ad(s)} = 0 ~,
\eea
and possess appropriate gauge freedoms; in the $s=0$ case one obtains:
\begin{subequations}
	\label{4.18}
	\begin{align}
		\d_L \bm{H}_{\a \ad} &= \bNabla_{\ad} L_{\a} - \Nabla_{\a} \bar{L}_{\ad} ~, \\
		\d_{\l, \L} \bm{\Psi}_{\a} &= \Nabla_{\a} \l + \L_\a ~, \qquad \bNabla_{\ad} \L_\a = 0~, \\
		\d_\chi \bm{G} &= \chi + \bar{\chi}~, \qquad \bNabla_{\ad} \chi = 0~,
	\end{align}
\end{subequations}
while for $s > 0$ we find:
\begin{subequations}
	\label{4.19}
	\begin{align}
		\d_L \bm{H}_{\a(s+1) \ad(s+1)} &= \bNabla_{(\ad_1} L_{\a(s+1) \ad_2 \dots \ad_{s+1})} - \Nabla_{(\a_1} \bar{L}_{\a_2 \dots \a_{s+1}) \ad(s+1)} ~, \\
		\d_{\eta,\l} \bm{\Psi}_{\a(s+1) \ad(s)} &= \bNabla_{(\ad_1} \eta_{\a(s+1) \ad_2 \dots \ad_{s})} + \Nabla_{(\a_1} \l_{\a_2 \dots \a_{s+1}) \ad(s)} ~, \\
		\d_\r \bm{G}_{\a(s) \ad(s)} &= \bNabla_{(\ad_1} \r_{\a(s) \ad_2 \dots \ad_{s})} - \Nabla_{(\a_1} \bar{\r}_{\a_2 \dots \a_{s}) \ad(s)}	~,
	\end{align}
\end{subequations}
where all gauge parameters are primary and complex unconstrained, unless otherwise stated. It should be noted that the component content of each $\N=1$ multiplet presented above was first reported in \cite{KMT}.

To conclude this section, we now reduce the action for $\U_{\a(s) \ad(s)}$ 
\begin{align}
	S^{(s,s)} = (-1)^s \int \rd^4x \rd^4 \q \, \cE\, \mathfrak{W}^{\a(2s+2)} (\U) \mathfrak{W}_{\a(2s+2)} (\U) + \text{c.c.}
\end{align}
from $\N=2$ to $\N=1$ conformal superspace (see also \cite{KR2021}). 
Omitting the technical details of the reduction, one arrives at the following:
\begin{align}
	\label{N=1actions}
	S^{(s,s)} &= (-1)^s \int \rd^4x \rd^2 \q \, \bm{\cE}\, \Big \{ 2 (2s+2)(2s+3) \bm{\mathfrak{W}}^{\a(2s+1)} (\bm{G}) \bm{\mathfrak{W}}_{\a(2s+1)} (\bm{G}) \non \\
	& \quad + 2 \ri \, \bm{\hat{\mathfrak{W}}}^{\a(2s+2)} (\bm \Psi) \bm{\check{\mathfrak{W}}}_{\a(2s+2)} (\bar{\bm \Psi}) - \hf \bm{\mathfrak{W}}^{\a(2s+3)} (\bm{H}) \bm{\mathfrak{W}}_{\a(2s+3)} (\bm{H}) \Big \} + \text{c.c.} 
\end{align}
where $\bm{\cE}$ denotes the $\N=1$ chiral integration measure and we have introduced the $\N=1$ higher-spin Weyl tensors:
\begin{subequations}
\begin{align}
	\bm{\hat{\mathfrak{W}}}_{\a(m+n+1)}(\bm{\U}) &= - \frac 14 \bNabla^2 \Nabla_{(\a_1}{}^{\b_1} \dots \Nabla_{\a_n}{}^{\b_n} \Nabla_{\a_{n+1}} \bm{\U}_{\a_{n+2} \dots \a_{m+n+1}) \b(n)} ~, \\
	\bm{\check{\mathfrak{W}}}_{\a(m+n+1)}(\bar{\bm{\U}}) &= - \frac 14 \bNabla^2 \Nabla_{(\a_1}{}^{\b_1} \dots \Nabla_{\a_m}{}^{\b_m} \Nabla_{\a_{m+1}} \bar{\bm{\U}}_{\a_{m+2} \dots \a_{m+n+1}) \b(m)} ~,
\end{align}
\end{subequations}
which are primary and gauge-invariant for any $\N=1$ SCHS gauge prepotential $\bm{\U}_{\a(m) \ad(n)}$, see \cite{KMT,KP19-2} for the technical details. We emphasise that in the case of a real prepotential, $m = n = s$, the field strengths coincide; $\bm{\hat{\mathfrak{W}}}_{\a(2s+1)}(\bm{\U}) = \bm{\check{\mathfrak{W}}}_{\a(2s+1)}(\bar{\bm{\U}}) \equiv \bm{{\mathfrak{W}}}_{\a(2s+1)}(\bm{\U})$.


\section{$\N > 2$ superconformal higher-spin gauge theories} \label{section4}

In the above sections we have performed a systematic study of $\N=2$ superconformal higher-spin theories in conformally-flat backgrounds. This resulted in the construction of new families of superconformal higher-spin multiplets and their corresponding gauge-invariant actions. Here we extend these results to the $\N > 2$ case.

The  conformal higher-spin $\cN=2$ supercurrents and their dual gauge prepotential were introduced in the previous sections for an arbitrary conformal supergravity background. 
Beyond $\cN=2$,  off-shell formulations  were constructed for $\cN=3$
\cite{vanMuiden:2017qsh, Hegde:2018mxv, Hegde:2021rte} and $\cN=4$ 
\cite{BdRdW, Butter:2016mtk, Butter:2019edc} conformal supergravity theories. 
The formalism of $\cN=4$ conformal superspace was briefly described in 
\cite{Butter:2019edc}, while $\cN=3$ conformal superspace has not yet been presented in the literature. Since currently there is no universal
 conformal superspace 
 formulation for $\cN$-extended conformal supergravity in four dimensions, in contrast to the three-dimensional case \cite{BKNT-M}, our analysis below will be restricted to conformally flat backgrounds.\footnote{In the component approach,  
the gauging of the $\cN$-extended superconformal group in four dimensions was studied for the first time in 
\cite{Ferrara:1977ij}, see also \cite{FT} for a review.}


\subsection{$\N > 2$ conformal superspace: conformally flat case}

We consider a conformally-flat $\N > 2$ superspace $\cM^{4|4\N}$, parametrised by local coordinates 
$z^{M} = (x^{m},\theta^{\m}_\imath,\bar \theta_{\dot{\mu}}^\imath)$, where 
$\imath = \underline{1}, \dots, \underline{\cN}$.
Its structure group is chosen to be the $\N$-extended superconformal group. The latter is spanned by  the Lorentz $M_{ab}$, translation $P_A=(P_a, Q_\a^I ,\bar Q^\ad_I)$, dilatation $\mathbb{D}$,  R-symmetry
$Y$ and $J^{I}{}_J$, and the special conformal $K^A=(K^a, S^\a_I ,\bar S_\ad^I)$ generators. Its geometry is encoded within the conformally covariant derivatives $\nabla_A = (\nabla_a,\nabla_\a^I,\bar{\nabla}_I^\ad)$, which take the form:
\begin{align}
	\nabla_A &= E_A - \hf \Omega_A{}^{ab} M_{ab} - \Phi_A{}^I{}_J J^{J}{}_I - \ri \Phi_A Y
	- B_A \mathbb{D} - \frak{F}_{AB} K^B \eol
	&= E_A - \Omega_A{}^{\b\g} M_{\b\g} - \bar{\Omega}_A{}^{\bd\gd} \bar{M}_{\bd\gd}
	- \Phi_A{}^I{}_J J^{J}{}_I - \ri \Phi_A Y - B_A \mathbb{D} - \frak{F}_{A B} K^B ~.
\end{align}

The algebraic relations obeyed by the generators of the $\N$-extended superconformal algebra take the same functional form as in the $\N=2$ case \eqref{N=2ConfAlg} with some minor exceptions. In particular, one should recall that, for $\N > 2$, the $\sSU(\N)_R$ generator $J^I{}_J$ acts on isospinors as follows
\bea
J^{I}{}_J \chi^{K} = - \d_J^K \chi^I + \frac{1}{\N} \d_J^I \chi^K ~.
\eea
Additionally, the $\sU(1)_R$ charges of the spinor covariant derivatives and special conformal generators now take the form:
\begin{align}
[Y, \nabla_\a^I] &= \frac{4 - \N}{\N} \nabla_\a^I ~, \qquad [Y, \bar{\nabla}^\ad_I] = \frac{\N - 4}{\N} \bar{\nabla}^\ad_I ~, \\
[Y, S^\a_I] &= \frac{\N-4}{\N} S^\a_I ~, \qquad [Y, \bar{S}_\ad^I] = \frac{4 - \N}{\N} \bar{S}_\ad^I ~.
\end{align}
The non-vanishing anti-commutation relations between 
spinor covariant derivatives and special conformal generators now take the form:
\begin{subequations}
\begin{align}
\{ S^\a_I , \nabla_\b^J \} &= \d^J_I \d^\a_\b \Big( 2 \mathbb{D} - Y \Big) - 4 \d^J_I M^\a{}_\b + 4 \d^\a_\b J^J{}_I ~, \quad \\
\{ \bar{S}^I_\ad , \bar{\nabla}^\bd_J \} &= \d^I_J \d^\bd_\ad \Big (2 \mathbb{D} + Y \Big )
+ 4 \d^I_J \bar{M}_\ad{}^\bd - 4 \d_\ad^\bd J^I{}_J ~.
\end{align}
\end{subequations}
It should be noted that, for $\N = 4$, the $\sU(1)_R$ generator becomes a central charge commuting with all elements of the superconformal algebra, see \cite{FT} 
for a detailed discussion. 

In general, the graded commutators $[\nabla_A, \nabla_B\}$ should be constrained in order to be expressed solely in terms of a single primary\footnote{A tensor superfield $\Psi$ with suppressed indices is said to be primary if it satisfies $K^B \Psi = 0$.} tensor superfield, the $\cN$-extended super-Weyl tensor $W$ (with suppressed indices) and its covariant derivatives. In this section, we will restrict our attention to so-called conformally-flat backgrounds, which are characterised by the constraint $W=0$. As a result, the only non-vanishing sector of $[\nabla_A, \nabla_B\}$ is
\bea
\{ \nabla_\a^I , \bar{\nabla}^{\ad}_J\} = - 2 \ri \d^I_J \nabla_\a{}^{\ad} ~.
\eea


\subsection{Conformal supercurrent multiplets and gauge prepotentials}

Let $m$ and $n$ be positive integers. A primary tensor superfield $\hat{J}^{\a(m) \ad(n)}$ defined on the background superspace will be called a conformal supercurrent if it obeys
\begin{subequations}
	\label{newSC1}
	\bea
	\nabla_\b^I \hat{J}^{\b \a(m-1) \ad(n)} &=& 0 \quad \Longrightarrow \quad \nabla^{IJ} \hat{J}^{\a(m) \ad(n)} = 0 ~, \\
	\bar{\nabla}_{\bd I} \hat{J}^{\a(m) \bd \ad(n-1)} &=& 0 \quad \Longrightarrow \quad \bar{\nabla}_{IJ} \hat{J}^{\a(m) \ad(n)} = 0 ~,
	\eea
\end{subequations}
where we have denoted
\bea
\nabla^{IJ} = \nabla^{\a(I} \nabla_\a^{ J)} ~, \qquad
\bar{\nabla}_{IJ} = \bar{\nabla}_{\ad (I} \bar{\nabla}^{\ad}_{J)} 
~.
\eea
These constraints uniquely fix the superconformal properties of $J^{\a(m) \ad(n)}$
\bea
\mathbb{D} \hat{J}^{\a(m) \ad(n)} = \hf (m+n+4) \hat{J}^{\a(m) \ad(n)} ~, \quad Y \hat{J}^{\a(m) \ad(n)} = - (m-n) \hat{J}^{\a(m) \ad(n)} ~.
\eea
When $n = 0$, the constraints 
\eqref{newSC1}
should be replaced with
\begin{subequations}
	\label{newSC2}
	\bea
	\nabla_\b^I \hat{J}^{\b \a(m-1)} &=& 0 \quad \Longrightarrow \quad \nabla^{IJ} \hat{J}^{\a(m) } = 0 ~, \\
	\bar{\nabla}_{IJ} \hat{J}^{\a(m)} &=& 0 ~.
	\eea
\end{subequations}
Their consistency with the superconformal algebra implies:
\bea
\mathbb{D} \hat{J}^{\a(m)} = \hf (m + 4) \hat{J}^{\a(m)} ~, \quad Y \hat{J}^{\a(m)} = - m \hat{J}^{\a(m)} ~.
\eea
Finally, for the special case $m=0$ , we replace 
\eqref{newSC2}
with
	\bea\label{newSC33}
	\nabla^{IJ} \hat{J} = 0 ~, \quad \bar{\nabla}_{IJ} \hat{J} = 0 ~,
	\eea
which imply:
\bea
\mathbb{D} \hat{J} = 2 \hat{J} ~, \quad Y \hat{J}= 0 ~.
\eea

In $\cN$-extended Minkowski superspace ${\mathbb M}^{4|4\cN}$, the conformal supercurrents $\hat{J}^{\a(n) \ad(n)}$ with $n>0$ were introduced in \cite{HST}. 
Certain  primary supermultiplets ${\mathbb M}^{4|4\cN}$ of the types
\eqref{newSC1}, \eqref{newSC2} and \eqref{newSC33} were also defined in \cite{Park}.

We now consider the Noether coupling 
\bea
\label{newNC}
S_{\text{NC}} = \int \rd^4x \rd^{2\N} \q \rd^{2\N} \bar\q \, \hat{E}\, \hat{\U}_{\a(m) \ad(n)} \hat{J}^{\a(m) \ad(n)}  ~,
\eea
where $\hat{E}$ is the full superspace measure and $\hat{\U}_{\a(m) \ad(n)}$ is the gauge prepotential dual to $\hat{J}^{\a(m) \ad(n)}$. For $m,n \geq 1$, \eqref{newNC} is inert under the transformations
\begin{subequations}
\label{newgt}
\bea
\label{newgt1}
\d_{\hat{\z},\hat{\l}} \hat{\U}_{\a(m) \ad(n)} = \nabla_{(\a_1}^I \hat{\z}_{\a_2 \dots \a_m) \ad(n) I} + \bar{\nabla}_{(\ad_1 I} \hat{\l}_{\a(m) \ad_2 \dots \ad_n)}{}^I~.
\eea
In the special case $n=0$, the above gauge transformation should be replaced with
\bea
\label{newgt2}
\d_{\hat{\z},\hat{\o}} \hat{\U}_{\a(m)} = \nabla_{(\a_1}^I \hat{\z}_{\a_2 \dots \a_m) I} + \bar{\nabla}_{IJ} \hat{\o}_{\a(m)}{}^{IJ}~.
\eea
Finally, setting $m=n=0$, this should be replaced with
\bea
\label{newgt3}
\d_{\hat{\o}} \hat{\U} = \bar{\nabla}_{IJ} \hat{\o}^{IJ} + \text{c.c.}
\eea
\end{subequations}
Requiring both the prepotentials $\hat{\U}_{\a(m) \ad(n)}$, $m,n \geq 0$, and corresponding gauge parameters to be primary uniquely fixes the dimension and $\sU(1)_R$ charge of the former:
\begin{subequations}
\label{SCprop}
\begin{align}
\mathbb{D} \hat{\U}_{\a(m) \ad(n)} &= - \hf (m + n + 4(\N - 1)) \hat{\U}_{\a(m) \ad(n)} ~, \\
\quad Y \hat{\U}_{\a(m) \ad(n)} &= (m-n) \hat{\U}_{\a(m) \ad(n)} ~.
\end{align}
\end{subequations}


\subsection{Superconformal gauge-invariant models}

In this subsection we present a gauge-invariant action for each gauge prepotential derived in the previous subsection. From the superfield $\hat{\U}_{\a(m) \ad(n)}$, $m,n \geq 0$, we construct the higher-derivative descendants
\begin{subequations}
	\label{newlinSW}
	\begin{align}
	\hat{\mathfrak{W}}_{\a(m+n+\N)} (\hat{\U}) &= \bar{\nabla}^{2 \N} \nabla_{(\a_1}{}^{\bd_1} \dots \nabla_{\a_n}{}^{\bd_n} \nabla_{\a_{n+1} \dots \a_{n+\N}} \hat{\U}_{\a_{n+\N+1} \dots \a_{m+n+\N}) \bd(n)} ~, \\
	\check{\mathfrak{W}}_{\a(m+n+\N)} (\hat{\bar{\U}}) &= \bar{\nabla}^{2\N} \nabla_{(\a_1}{}^{\bd_1} \dots \nabla_{\a_m}{}^{\bd_m} \nabla_{\a_{m+1} \dots \a_{m+\N}} \hat{\bar{\U}}_{\a_{m+\N+1} \dots \a_{m+n+\N}) \bd(m)} ~.
	\end{align}
\end{subequations}
Here we have introduced the $\N$th order operators\footnote{For simplicity, we do not include any numerical factors in \eqref{4.177} and \eqref{4.188}.}
\bea
\nabla_{\a(\N)}= \ve_{I_1 \dots I_\N} \nabla_{(\a_1}^{I_1} \dots \nabla_{\a_\N)}^{I_\N} ~, \quad \bar{\nabla}^{\ad(\N)} = \ve^{I_1 \dots I_\N} \bar{\nabla}^{(\ad_1}_{I_1} \dots \bar{\nabla}^{\ad_\N)}_{I_\N}~,
\label{4.177}
\eea
where $\ve^{1 \dots \N} = \ve_{1 \dots \N}  = 1$, as well as the chiral projection operator 
\bea
\bar{\nabla}^{2 \N}  = \bar{\nabla}_{\ad(\N)} \bar{\nabla}^{\ad(\N)} ~.
\label{4.188}
\eea
In the case of a Minkowski background, the field strengths \eqref{newlinSW}  coincide with those introduced in \cite{GGRS}. 

The chiral field strengths \eqref{newlinSW} have the following dimensions:
\begin{subequations}
	\bea
	\mathbb{D} \hat{\mathfrak{W}}_{\a(m+n+\N)} (\U) &=&  \hf (n-m - \N + 4) \hat{\mathfrak{W}}_{\a(m+n+\N)} (\U) ~, \\
	\mathbb{D} \check{\mathfrak{W}}_{\a(m+n+\N)} (\bar \U) &=& \hf (m-n - \N + 4) \check{\mathfrak{W}}_{\a(m+n+\N)}  (\bar \U) ~,
	\eea
\end{subequations}
and may be shown to be primary
\bea
K^B \hat{\mathfrak{W}}_{\a(m+n+\N)} (\U) = 0 ~, \quad K^B \check{\mathfrak{W}}_{\a(m+n+\N)} (\bar \U) = 0 ~,
\eea
and are invariant under the gauge transformations \eqref{newgt}.

These properties imply that the action
\bea
\label{newSCHSaction}
S^{(m,n)} = \ri^{m+n} \int \rd^4x \rd^{2\N} \q \, \hat{\cE} \, \hat{\mathfrak{W}}^{\a(m+n+\N)} (\U) \check{\mathfrak{W}}_{\a(m+n+\N)} (\bar \U) + \text{c.c.}
\eea
is locally superconformal and gauge-invariant. Here $\hat{\cE}$ is the chiral integration measure. 
The overall factor of $\ri^{m+n}$ in \eqref{SCHSaction1} has been chosen due to the identity
\bea
\ri^{m+n+1} \int \rd^4x \rd^4 \q \, \hat{\cE} \, \hat{\mathfrak{W}}^{\a(m+n+\N)} (\U) \check{\mathfrak{W}}_{\a(m+n+\N)} (\bar \U) + \text{c.c.} = 0~,
\eea
which holds up to a total derivative. 


\section{Conclusion} \label{section5}

In this work we have, for the first time, introduced the $\cN=2$ superconformal higher-spin gauge prepotentials $\U_{\a(m) \ad(n)}$ in conformal superspace, with $m \geq n \geq 0$, and derived their gauge-invariant actions in arbitrary conformally flat backgrounds.\footnote{It should be emphasised once more that the actions \eqref{SCHSaction1}, \eqref{SCHSaction2} and \eqref{SCHSaction3} 
are locally superconformal in any curved background. But they are gauge invariant only if the background super-Weyl tensor vanishes.} These results have also been generalised to $\cN>2$. In both $\cN=2$ and $\cN>2$ cases,  the  
superconformal higher-spin gauge prepotentials originate as the duals of 
 the superconformal higher-spin multiplets of conserved currents, which have been described in our paper.

Our work provides the stepping stones to address several interesting problems.
First of all, it would be interesting to extend some of our $\cN=2$ superconformal gauge models to non-conformally flat supergravity backgrounds.\footnote{The construction of gauge-invariant models for conformal higher-spin (super)fields in curved (super)gravity backgrounds has been an active  area of research \cite{NT, GrigorievT,KMT, BeccariaT, Manvelyan, KP, KP19-2, KPR1,KPR2}.} In particular, there must exist a consistent deformation of  the linearised action 
for $\cN=2$ conformal supergravity \eqref{SCHSaction3} to arbitrary Bach-flat backgrounds described by eq. \eqref{A.15}.

Our  $\cN=2$ superconformal higher-spin models have already been demonstrated to be
$\sU(1)$ duality invariant \cite{KR2021}. It would be interesting to analyse whether their 
$\cN>2$ cousins constructed in section \ref{section4} share this remarkable property. 

It is of interest to study the problem of 
using Vasiliev's unfolded dynamics formalism \cite{Vasiliev:1988sa}
 in order to reconstruct our superspace formulations directly from the component ones along the lines of the approach of \cite{MV}.  The latter can be applied to the construction of conformal higher-spin currents of \cite{GV} uplifted to arbitrary conformally flat background. 
  These issues will be addressed elsewhere.

In the remainder of this section we will concentrate, in some more detail, 
on several implications and natural extensions of our results. 


\subsection{$\cN=2$ superconformal higher-spin gauge theories in AdS$_4$}

Fundamental to our analysis is the formalism of conformal superspace, which trivialises calculations in conformally flat backgrounds. For the purpose of applications, however, it is often useful to work with Lorentz covariant derivatives as opposed to their conformally covariant counterparts. The process of translating results from conformal to $\sU(2)$ superspace is known as degauging \cite{ButterN=2} and while the general procedure is well-known, it is often highly non-trivial to perform on generic curved backgrounds. Such computations are greatly simplified when we restrict the geometry by turning off several torsion superfields.

To this end, we now degauge and reiterate our main results in the $\mathcal{N}=2$ AdS superspace.\footnote{The conformal flatness of  $\mathcal{N}$-extended  AdS superspace in four dimensions was established in \cite{BILS}.}
Recall that its geometry is encoded within the covariant derivatives $\cD_A = (\cD_a, \cD_\a^i, \bar \cD^\ad_i)$, which obey the algebra (see, e.g. \cite{KLRT-M1, KT-M-ads}):
\be
\{ \cD_\a^i , \cD_\b^j \} = 4 S^{ij} M_{\a \b} + 2 \ve_{\a \b} \ve^{i j} S^{kl} J_{kl} ~, \quad \{ \cD_\a^i , \bar \cD^\bd_j \} = - 2 \ri \d_j^i \cD_{\a}{}^{\bd} ~,
\ee
where ${S}^{ij} $ is a  covariantly constant
  iso-triplet, ${S}^{ji} = { S}^{ij}$, satisfying  the integrability condition $[S, S^\dagger ]=0$, 
  with $S= (S^i{}_j)$.\footnote{The integrability condition implies that $S^{ij}$ can be chosen to be real,   
 $\overline{ {S}^{ij}} = {S}_{ij} =\ve_{ik}\ve_{jl}{ S}^{kl}$. However we will not impose the reality condition.}

We first derived the $\N=2$ conformal supercurrents, which are described by the tensor superfields $J^{\a(m) \ad(n)}$ with $m,n \geq 0$. For $m,n \geq 1$, they obey:
\begin{subequations}
\label{AdSSC1}
\bea
\cD_\b^i J^{\b \a(m-1) \ad(n)} &=& 0 \quad \Longrightarrow \quad (\cD^{ij} + 2(m+2) S^{ij}) J^{\a(m) \ad(n)} = 0 ~, \\
\bar{\cD}_\bd^i J^{\a(m) \bd \ad(n-1)} &=& 0 \quad \Longrightarrow \quad (\bar{\cD}^{ij} + 2(n+2) \bar{S}^{ij}) J^{\a(m) \ad(n)} = 0 ~,
\eea
\end{subequations}
where we have defined $\cD^{ij} = \cD^{\a(i} \cD_{\a}^{j)}$ and $\bar \cD^{ij} = \bar{\cD}_{\ad}^{(i} \bar \cD^{\ad j)}$. When $n = 0$ \eqref{AdSSC1} should be replaced with
\begin{subequations}
\label{AdSSC2}
\bea
\cD_\b^i J^{\b \a(m-1)} &=& 0 \quad \Longrightarrow \quad (\cD^{ij} + 2(m+2) S^{ij}) J^{\a(m) } = 0 ~, \\
(\bar{\cD}^{ij} + 4 \bar{S}^{ij}) J^{\a(m)} &=& 0 ~.
\eea
\end{subequations}
Finally, for $m=n=0$ we exchange \eqref{AdSSC2} with
\bea
({\cD}^{ij} + 4 {S}^{ij}) J = 0~, \quad (\bar{\cD}^{ij} + 4 \bar{S}^{ij}) J = 0~.
\eea

The supercurrents described above are dual, via the Noether coupling \eqref{NC}, to gauge prepotentials $\U_{\a(m) \a(n)}$, with $m,n \geq 0$. For $m,n \geq 1$, they are defined modulo the gauge transformations
\bea
\label{AdSSCHSgt1}
\d_{\z,\l} \U_{\a(m) \ad(n)} = \cD_{(\a_1}^i \z_{\a_2 \dots \a_m) \ad(n) i} + \bar{\cD}_{(\ad_1}^i \l_{\a(m) \ad_2 \dots \ad_n) i} ~.
\eea
Here the parameters $\z_{\a(m-1) \ad(n) i}$ and $\l_{\a(m) \ad(n-1) i }$ are complex unconstrained. When $n=0$ the gauge transformation law becomes
\bea
\label{AdSSCHSgt2}
\d_{\z,\o} \U_{\a(m)} = \cD_{(\a_1}^i \z_{\a_2 \dots \a_m) i} + (\bar{\cD}^{ij} + 4 \bar{S}^{ij}) \o_{\a(m) ij} ~.
\eea
Here $\z_{\a(m-1) i}$ and $\o_{\a(m) ij }$ are complex unconstrained. Finally, for $m=0$ the gauge transformation law is
\bea
\label{AdSSCHSgt3}
\d_{\o} \U = (\bar \cD^{ij} + 4 \bar S^{ij}) \o_{ij} + \text{c.c.}~,
\eea
where $\o_{ij}$ is complex unconstrained. This prepotential describes the conformal supergravity multiplet. 

Associated with $\U_{\a(m) \ad(n)}$ (and its conjugate $\bar \U_{\a(n) \ad(m)}$) are the field strengths
\begin{subequations}
\label{AdSLinSW}
\bea
\hat{\mathfrak{W}}_{\a(m+n+2)} (\U) &=& \frac{1}{48} (\bar \cD^{ij} + 4 \bar S^{ij})\bar \cD_{ij} \cD_{(\a_1}{}^{\bd_1} \dots \cD_{\a_n}{}^{\bd_n} \non \\
&& \qquad \qquad  \qquad \qquad \times \cD_{\a_{n+1} \a_{n+2}} \U_{\a_{n+3} \dots \a_{m+n+2}) \bd(n)} ~, \\
\check{\mathfrak{W}}_{\a(m+n+2)} (\bar \U) &=& \frac{1}{48} (\bar \cD^{ij} + 4 \bar S^{ij}) \bar \cD_{ij} \cD_{(\a_1}{}^{\bd_1} \dots \cD_{\a_m}{}^{\bd_m} \non \\
&& \qquad \qquad  \qquad \qquad \times \cD_{\a_{m+1} \a_{m+2}} \bar \U_{\a_{m+3} \dots \a_{m+n+2}) \bd(m)} ~,
\eea
\end{subequations}
which are the linearised higher-spin super-Weyl tensors. As a final step, they may be employed to construct the superconformal gauge-invariant actions
\bea
\label{AdSACtion}
S^{(m,n)} = \ri^{m+n} \int \rd^4x \rd^4 \q \, \cE\, \hat{\mathfrak{W}}^{\a(m+n+2)} (\U) \check{\mathfrak{W}}_{\a(m+n+2)} (\bar \U) + \text{c.c.}
\eea


\subsection{$\cN=2$ superconformal higher-spin models as induced actions}

An interesting problem is to derive the $\cN=2$ superconformal higher-spin models proposed 
in section \ref{section3} as induced actions, in the spirit of the non-supersymmetric studies carried out in \cite{Tseytlin,Segal,BJM,Bonezzi,BeccariaT}. 
One possible approach is to couple an off-shell superconformal hypermultiplet to 
background gauge prepotentials   $\U_{\a(s) \ad(s) } =\bar \U_{\a(s) \ad(s) } $ using an action of the form
\bea
S[q,\bar q;\U]= S_{\rm hyper} [q , \bar q] +\sum_{s=0}^{\infty}
\int \rd^4x\rd^4\theta\rd^4\bar\theta\, E\, \U_{\a(s) \ad(s) } J^{\a(s) \ad(s)} ~,
\label{4.1}
\eea
 where  $S_{\rm hyper} $ denotes an off-shell action for the superconformal massless hypermultiplet $q$, and $J^{\a(s) \ad(s)} $ conserved higher-spin supercurrents in the model $ S_{\rm hyper} $. Then it is natural to consider 
the generating functional for correlation functions of these conserved higher spin 
supercurrents defined by
\bea
\re^{ \ri \,\G[\U] } = \int \cD q \cD \bar q \,\re^{ \ri \,S[q, \bar q; \U]}~.
\label{4.2}
\eea
Integrating out the hypermultiplet superfields $q$ and $\bar q$ and computing the logarithmically divergent part of the effective action, one is expected to end up with 
the $\cN=2$ superconformal higher-spin models described
in section \ref{section3}.\footnote{This proposal is analogous to the one put forward in the $\cN=1$ superconformal case \cite{KMT}.}
There exist two powerful superspace approaches which offer off-shell formulations for the charged hypermultiplet: the harmonic superspace \cite{GIKOS,GIOS} and the projective superspace \cite{KLR,LR1,LR2}.\footnote{See \cite{K2010} for a brief review of these approaches and their relationship.} 
Both formulations can be used to do the path integral \eqref{4.2}.
  
It remains to comment on the higher-spin supercurrents in \eqref{4.1}.
For this we consider an on-shell hypermultiplet, which is described by the primary isospinor $q^i$ (and its conjugate $\bar q_i$) subject to the constraints
\bea
\nabla_{\a}^{(i} q^{j)} = 0~, \quad \bar \nabla_{\ad}^{(i} q^{j)} = 0~.
\eea
These are consistent with the superconformal algebra provided
\bea
\mathbb{D} q^i = q^i~, \quad Y q^i = 0~.
\eea
The hypermultiplet supercurrent is 
\bea 
J= -\hf q^i \bar q_i~,
\eea
see \cite{KT} for more details. 
The hypermultiplet higher-spin supercurrents $J^{\a(s) \ad(s)}$, with $s>0$, 
are uniquely obtained as the following composites of $q^i$ and $\bar{q_i}$:
\bea
\label{HMSC}
J^{\a(s) \ad(s)} &=&\ri^{s} \sum^s_{k=0} (-1)^k {s \choose k}^2 \nabla^{(\a_1 (\ad_1} \dots \nabla^{\a_k \ad_k} q^i \nabla^{\a_{k+1} \ad_{k+1}} \dots \nabla^{\a_s) \ad_s)} \bar{q}_i \non \\
&&- \frac{\ri^{s+1} } 8 \sum_{k=0}^{s-1} (-1)^k {s \choose k} {s \choose k+1} \non \\
&&\qquad  \times 
\bigg\{ \nabla^{(\a_1 (\ad_1} \dots \nabla^{\a_k \ad_k} \nabla^{\a_{k+1} i} q_i \nabla^{\a_{k+2} \ad_{k+1}} \dots \nabla^{\a_s) \ad_{s-1}} \bar \nabla^{\ad_s) j} \bar q_j \non \\
&&\qquad \quad - \nabla^{(\a_1 (\ad_1} \dots \nabla^{\a_k \ad_k} \bar \nabla^{\ad_{k+1} i} q_i \nabla^{\a_{k+1} \ad_{k+2}} \dots \nabla^{\a_{s-1} \ad_{s})} \nabla^{\a_s) j}
\bar q_j
\bigg\} ~.
\eea
It is instructive to verify that \eqref{HMSC} does indeed satisfy the conservation equations \eqref{SC1} in any conformally flat background.

To connect our results to the existing literature on superconformal higher-spin multiplets, it is instructive to comment on the reduction of \eqref{4.1} to $\N=1$ superspace. First, we recall that at the $\N=1$ level the on-shell hypermultiplet is described by two free, massless chiral scalars:
\begin{subequations}
	\begin{align}
		\Phi_{(+)} = \bar{q}_{\underline{1}} | ~, \qquad \bNabla_\ad \Phi_{(+)} = 0 ~, \qquad \Nabla^2 \Phi_{(+)} = 0 ~,
		\\ \Phi_{(-)} = q^{\underline{2}} | ~, \qquad \bNabla_\ad \Phi_{(-)} = 0 ~, \qquad \Nabla^2 \Phi_{(-)} = 0 ~.
	\end{align}
\end{subequations}
This result, combined with definitions \eqref{2.16}, allows one to prove that the $\N=1$ supercurrent multiplets encoded within \eqref{HMSC} coincide with those of \cite{KMT}. This, in addition to the results of section \ref{section3.4}, allows us to reduce \eqref{4.1} to $\N=1$ superspace
\begin{align}
	S[\Phi_{(\pm)}&;\bm{H},\Psi,\bm{G}] = S_{\rm hyper} [\Phi_{(\pm)}] +\sum_{s=0}^{\infty}
	\int \rd^4x\rd^2\theta\rd^2\bar\theta\, E \, \Big \{ \frac 1 4 \bm{H}_{\a(s+1) \ad(s+1) } j^{\a(s+1) \ad(s+1)}  \non \\
	&+ \hf \bm{\Psi}_{\a(s+1) \ad(s)} j^{\a(s+1) \ad(s)} - \hf \bar{\bm{\Psi}}_{\a(s) \ad(s+1)} \bar{j}^{\a(s) \ad(s+1)} + \bm{G}_{\a(s) \ad(s)} j^{\a(s) \ad(s)} \Big \} ~.
\end{align}
Upon integrating out the chiral superfields $\Phi_{(\pm)}$ and extracting the logarithmically divergent part of the effective action, we expect to arrive at the models described by \eqref{N=1actions}.

\subsection{$\cN=2$ superconformal vector and gravitino multiplets} 

All chiral field strengths $\hat{\mathfrak{W}}_{\a(k)}$ and $\check{\mathfrak{W}}_{\a(k)}$ constructed in section \ref{section3} carry at least two spinor indices, $k\geq 2$. The case $k=0$ correspond to the massless vector multiplet \cite{GSW}
which can be described in terms of the curved superspace analogue
of Mezincescu's prepotential \cite{Mezincescu} (see also \cite{HST}),  $V_{ij}=V_{ji}$. 
The latter is a primary unconstrained real $\sSU(2)$ triplet of dimension $-2$. 
The expression for $\mathfrak{W}$ in terms of $V_{ij}$ is
(see \cite{n2_sugra_tensor} for the derivation) 
\bea
\mathfrak{W} (V) = \frac{1}{4}\bar\nabla^4 \nabla^{ij} V_{ij}~,
\eea
and it defines a primary reduced  chiral superfield of dimension $+1$, 
\bea
\nabla^{ij} \mathfrak{W} = \bar \nabla^{ij} \bar{\mathfrak{W}}~.
\eea
The field strength $\mathfrak W$ is invariant under gauge transformations of the form
\bea
\delta V^{ij} &= \nabla^{\alpha}{}_k \Lambda_\alpha{}^{kij}
+ \bar\nabla_{\dalpha}{}_k \bar\Lambda^\dalpha{}^{kij}, \qquad
\Lambda_\alpha{}^{kij} = \Lambda_\alpha{}^{(kij)}~,
\eea
with $ \Lambda_\alpha{}^{kij} $ being primary and unconstrained modulo the algebraic condition given.

It would be interesting to understand what is a gauge prepotential corresponding to the chiral field strengths $\hat{\mathfrak{W}}_{\a}$ and $\check{\mathfrak{W}}_{\a}$. 
These field strengths should correspond to a superconformal gravitino multiplet.

It should be pointed out that the multiplet of currents associated with Mezincescu's prepotential  is the so-called linear multiplet \cite{BS,SSW}, $J^{ij}$, which is a primary real iso-triplet of dimensions $+2$  constrained by 
\bea
\nabla^{(i}_\a J^{jk)} =  {\bar \nabla}^{(i}_\ad J^{jk)} = 0~.
\eea
in the hypermultiplet case $J^{ij}= \ri \bar q^{(i} q^{j)}$. Strictly speaking, 
the action \eqref{4.1} should include an additional term, 
$\int \rd^4x\rd^4\theta\rd^4\bar\theta\, E\,V_{ij} J^{ij}$.


\subsection{Massless higher-spin $\cN=2$ supermultiplets in AdS$_4$}

It is known that massless higher-spin gauge models can be realised in terms of a conformal prepotentials coupled to lower-spin compensators. In particular, this is true of the non-supersymmetric (Fang-)Fronsdal  actions in AdS$_4$ \cite{F,FF}, as well as  their $\cN=1$ supersymmetric counterparts \cite{KS94,BHK}. In the present work we have identified the $\cN=2$ superconformal higher-spin gauge prepotentials, which opens up 
the opportunity to construct off-shell massless  higher-spin $\cN=2$ supermultiplets in a manifestly 
$\sOSp (2|4) $ invariant setting.\footnote{In terms of $\cN=1$ superfields, the off-shell 
massless higher-spin $\cN=2$ supersymmetric gauge models in AdS$_4$ were constructed in \cite{GKS1,GKS2}.}  
So far, compensating superfields and gauge-invariant actions  have been found only for linearised $\cN=2$ supergravity \cite{ButterK}.
Nevertheless, our construction naturally lends itself to the description of on-shell massless higher-spin supermultiplets that should originate in the massless higher-spin supersymmetric models as the only gauge-invariant field strengths which survive on the mass shell. In the $\N=2$ AdS superspace, they are described by primary tensor superfields 
${\mathbb W}_{\a(k)}$, with $k \geq 1$, satisfying the constraints
\bea
\bar{\cD}_{\ad}^i {\mathbb W}_{\a(k)} &=& 0 ~, \quad
\cD^{\b i} \mathbb{W}_{\a(k-1) \b} = 0 ~.
\eea
These equations furnish a representation of the $\cN=2$ superconformal group, since they 
are the degauged form of the following equations in conformal superspace
\bea
\bar{\nabla}_{\ad}^i {\mathbb W}_{\a(k)} &=& 0 ~, \quad
\nabla^{\b i} \mathbb{W}_{\a(k-1) \b} = 0 ~.
\eea
These are conformally invariant provided the primary chiral superfield $\mathbb{W}_{\a(k)} $ is characterised by the dimension
\bea
\mathbb{D}\, \mathbb{W}_{\a(k)} =  \hf ( k + 2) \mathbb{W}_{\a(k)} ~.
\label{5.24}
\eea

It is worth considering in more detail the case of an even $k$.
Given a non-negative integer $s$, $\mathbb{W}_{\a(2s+2)} $ is constructed in terms of 
a real prepotential $\mathbb{H}_{\a(s) \ad(s)} $ according to 
\bea
\mathbb{W}_{\a(2s+2)} ({\mathbb H} ) &=& \frac{1}{48} (\bar \cD^{ij} + 4 \bar S^{ij})\bar \cD_{ij} \cD_{(\a_1}{}^{\bd_1} \dots \cD_{\a_s}{}^{\bd_s} \non \\
&& \qquad \qquad  \qquad \qquad \times \cD_{\a_{s+1} \a_{s+2}} {\mathbb H}_{\a_{s+3} \dots \a_{2s+2}) \bd_1 \dots \bd_s} ~,
\label{5.25}
\eea
which should be compared with \eqref{AdSLinSW}. This field strength is invariant under the gauge transformations
\bea
\d_\z {\mathbb H}_{\a(s) \ad(s)} = \cD_{(\a_1}^i \z_{\a_2 \dots \a_s) \ad(s) i} - \bar{\cD}_{(\ad_1}^i \bar{\z}_{\a(s) \ad_2 \dots \ad_s) i} ~.
\eea
This gauge symmetry is similar to \eqref{realPPgt}, which allows us to refer to 
${\mathbb H}_{\a(s) \ad(s)} $ as the conformal prepotential for a massless theory.\footnote{The massless theory also involves some compensating multiplets. They do not appear in the field strength \eqref{5.25}.} However, this terminology is not quite precise, since ${\mathbb H}_{\a(s) \ad(s)} $ is not a primary superfield.
This follows from the fact that the dimension of $\mathbb{W}_{\a(2s+2)} $ is $s+2$, 
eq. \eqref{5.24}, while the dimension of $ {\mathfrak{W}}_{\a(2s+2)} (\U) $
is equal to $+1$, eq. \eqref{3.100}, for  real $ \U_{\a(s) \ad(s)} $.

It is instructive to comment on the $\N=1$ superfield content of these supermultiplets. In particular, the independent components are
\begin{subequations}
\bea
\mathcal{F}_{\a(k)} &=& \mathbb{W}_{\a(k)} | ~, \\
\mathcal{G}_{\a(k+1)} &=& \cD_{(\a_1}^{\underline{2}} \mathbb{W}_{\a_2 \dots \a_{k+1})} | ~,
\eea
\end{subequations}
which prove to be  on-shell massless $\N=1$ conformal superfields
\begin{subequations}
\bea
\bar{\bm{\cD}}_{\ad} \mathcal{F}_{\a(k)} &=& 0 ~, \quad \bm{\cD}^{\b} \mathcal{F}_{\a(k-1) \b} = 0 ~, \\
\bar{\bm{\cD}}_{\ad} \mathcal{G}_{\a(k+1)} &=& 0 ~, \quad \bm{\cD}^{\b} \mathcal{G}_{\a(k) \b} = 0 ~,
\eea
\end{subequations}
where $\bm{\cD}_{A}$ are the covariant derivatives of the $\N=1$ AdS superspace, see e.g. \cite{BK}. These on-shell $\cN=1$ supermultiplets naturally occur in the $\cN=1$ massless higher-spin gauge theories \cite{KS94,BHKP}.
\\


\noindent
{\bf Acknowledgements:}\\
We thank the referees for their useful suggestions.
The work of SK is supported in part by the Australian 
Research Council, project No. DP200101944.
The work of ER is supported by the Hackett Postgraduate Scholarship UWA,
under the Australian Government Research Training Program.

\appendix

\section{$\mathcal{N}=2$ conformal superspace in four dimensions}\label{AppendixA}

This appendix reviews $\mathcal{N} = 2$ conformal superspace, a formulation for off-shell $\mathcal{N}=2$ conformal supergravity developed by Butter \cite{ButterN=2} and then reformulated in \cite{BN}. Our spinor conventions are those of \cite{BK}, which are similar
to \cite{WB}.

We consider a curved $\cN=2$ superspace $\mathcal{M}^{4|8}$
parametrised by local coordinates 
$z^{M} = (x^{m},\theta^{\m}_\imath,\bar \theta_{\dot{\mu}}^\imath)$, where $m = 0,1,2,3$, $\m = 1,2$ and $\imath = \underline{1}, \underline{2}$. 
The structure group is chosen to be $\sSU(2,2|2)$. The corresponding superalgebra is spanned by the Lorentz $M_{ab}$, translation $P_A=(P_a, Q_\a^i ,\bar Q^\ad_i)$, dilatation $\mathbb{D}$,  R-symmetry $Y$ and $J^{ij}$, and the special conformal $K^A=(K^a, S^\a_i ,\bar S_\ad^i)$ generators. The covariant derivatives $\nabla_A = (\nabla_a, \nabla_\alpha^i, \bar\nabla^\dalpha_i)$ then have the form
\begin{align}
\nabla_A &= E_A - \hf \Omega_A{}^{ab} M_{ab} - \Phi_A{}^{ij} J_{ij} - \ri \Phi_A Y
- B_A \mathbb{D} - \frak{F}_{A B} K^B \eol
&= E_A - \Omega_A{}^{\b\g} M_{\b\g} - \bar{\Omega}_A{}^{\bd\gd} \bar{M}_{\bd\gd}
- \Phi_A{}^{ij} J_{ij} - \ri \Phi_A Y - B_A \mathbb{D} - \frak{F}_{A B} K^B ~.
\label{A.1}
\end{align}
Here $E_A = E_A{}^M \pa_M$ is the supervielbein, $\Omega_A{}^{ab}$ the Lorentz connection,
and $\Phi_A{}^{ij}$ and $\Phi_A$ are the $\sSU(2)_R$ and $\sU(1)_R$ connections,
respectively. In addition, we have a dilatation connection $B_A$ and a special
superconformal connection $\frak F_{AB}$.

The Lorentz ($M_{ab}$) and $\sSU(2)_R$ ($J^{ij}$) generators are defined to act on Weyl spinors, vectors and isospinors in the following way:
\begin{subequations}
\begin{align}
M_{\a \b} \psi_\g &= \ve_{\g (\a} \psi_{\b)} ~, \quad \bar M_{\ad \bd} \bar \psi_\gd = \ve_{\gd (\ad} \psi_{\bd)} ~, \\
M_{ab} V_c &= 2 \eta_{c [a} V_{b]} ~, \quad J^{ij} \chi^k = \ve^{k(i} \chi^{j)} ~.
\end{align}
\end{subequations}
The $\sU(1)_R$ and dilatation generators obey:
\begin{subequations}
\label{N=2ConfAlg}
\begin{align}
[Y, \nabla_\a^i] &= \nabla_\a^i ~,\quad [Y, \bar\nabla^\ad_i] = - \bar\nabla^\ad_i~,  \non \\
[\mathbb{D}, \nabla_a] &= \nabla_a ~, \quad
[\mathbb{D}, \nabla_\a^i] = \hf \nabla_\a^i ~, \quad
[\mathbb{D}, \bar\nabla^\ad_i] = \hf \bar\nabla^\ad_i ~.
\end{align}
At the same time, the special superconformal generators $K^A = (K^a, S^\alpha_i, \bar S_\dalpha^i)$ carry opposite $\sU(1)_R$ and dilatation weight to $\nabla_A$:
\begin{align}
[Y, S^\a_i] &= - S^\a_i ~, \quad
[Y, \bar{S}^i_\ad] = \bar{S}^i_\ad~, \non \\
[\mathbb{D} , K_a] &= - K_a ~, \quad
[\mathbb{D}, S^\a_i] = - \hf S^\a_i ~, \quad
[\mathbb{D}, \bar{S}_\ad^i] = - \hf \bar{S}_\ad^i ~.
\end{align}
Among themselves, the generators $K^A$ obey the algebra
\begin{align}
\{ S^\a_i , \bar{S}^j_\ad \} &= 2 \ri \d^j_i K^\a{}_{\ad}~,
\end{align}
with all the other (anti-)commutators vanishing. 	Finally, the algebra of $K^A$ with $\nabla_B$ is given by
\begin{align}
[K_\aa, \nabla_\bb] &= -4 \ve_{\a\b} \ve_{\ad \bd} \mathbb{D} + 4 \ve_{\ad \bd} M_{\a \b} + 4 \ve_{\a \b} \bar M _{\ad \bd} ~,\non \\
\{ S^\a_i , \nabla_\b^j \} &= \d^j_i \d^\a_\b (2 \mathbb{D} - Y) - 4 \d^j_i M^\a{}_\b + 4 \d^\a_\b J_i{}^j ~,\non \\
\{ \bar{S}^i_\ad , \bar{\nabla}^\bd_j \} &= \d^i_j \d^\bd_\ad (2 \mathbb{D} + Y)
+ 4 \d^i_j \bar{M}_\ad{}^\bd - 4 \d_\ad^\bd J^i{}_j ~,\non \\
[K_\aa, \nabla_\b^j] &= - 2 \ri \ve_{\a \b} \bar{S}_\ad^j \ , \quad [K_\aa, \bar{\nabla}^\bd_j] = 
- 2\ri \d_\ad^\bd S_{\a j} ~, \non \\
[S^\a_i , \nabla_\bb] &= 2 \ri \d^\a_\b \bar{\nabla}_\bd^i \ , \quad [\bar{S}^i_\ad , \nabla_\bb ] = 
- 2 \ri \ve_{\ad \bd} \nabla_{\b i} \ ,
\end{align}
\end{subequations}
where all other graded commutators vanish.

Consider a tensor superfield $\J$ with suppressed indices. It is said to be primary if
\bea
K^B \Psi = 0 ~.
\eea
Additionally, its dimension $\D_\Psi$ and $\sU(1)_R$ charge $q_\Psi$ are defined as follows:
\bea
\mathbb{D} \Psi = \D_\Psi \Psi ~, \quad Y \Psi = q_\Psi \Psi ~.
\eea
Of particular importance are (covariantly) chiral superfields, which satisfy
\bea
\bar{\nabla}_{\ad}^i \J = 0~.
\eea
The consistency of this constraint with the superconformal algebra leads to highly non-trivial implications. In particular, it can carry no isospinor or dotted spinor indices, $\Psi = \Psi_{\a(m)}$, and its $\sU(1)_R$ charge and dimension are related as follows:
\bea
\label{chiralDimCharge}
q_\Psi = - 2 \D_\Psi ~.
\eea
Further, we note that for any primary tensor superfield $\Phi_{\a(m)}$ with the property $q_\Phi = - 2 \D_\Phi$, the following object 
\bea
\Psi_{\a(m)} = \bar{\nabla}^4 \Phi_{\a(m)} \equiv \frac{1}{48} \bar{\nabla}^{ij} \bar{\nabla}_{ij} \Phi_{\a(m)}
\eea
is both chiral and primary \cite{KTM08}. 
It is important to note that not all chiral primary superfields take this form, see e.g. \eqref{linSW1}. 

In \cite{ButterN=2}, it was shown that, in order to reproduce the component structure of conformal supergravity, certain constraints must be imposed on the graded commutators $[\nabla_A , \nabla_B \}$. In particular, they must be expressed solely in terms of the super-Weyl tensor, $W_{\a(2)}$, and its covariant derivatives. This superfield is primary, chiral and carries dimension $1$,
\bea
K^B W_{\a \b}  = 0 ~, \quad \bar{\nabla}_\ad W_{\a \b} = 0 ~, \quad \mathbb{D} W_{\a \b} = W_{\a \b}.
\eea
The solution to the aforementioned constraints are given by
\begin{subequations}\label{CSGAlgebra}
\begin{align}
\{ \nabla_\a^i , \nabla_\b^j \} &= 2 \ve^{ij} \ve_{\a\b} \bar{W}_{\gd\dd} \bar{M}^{\gd\dd} + \hf \ve^{ij} \ve_{\a\b} \bar{\nabla}_{\gd k} \bar{W}^{\gd\dd} \bar{S}^k_\dd - \hf \ve^{ij} \ve_{\a\b} \nabla_{\g\dd} \bar{W}^\dd{}_\gd K^{\g \gd}~, \\
\{ \nabla_\a^i , \bar{\nabla}^\bd_j \} &= - 2 \ri \d_j^i \nabla_\a{}^\bd~, \\
[\nabla_{\a\ad} , \nabla_\b^i ] &= - \ri \ve_{\a\b} \bar{W}_{\ad\bd} \bar{\nabla}^{\bd i} - \frac{\ri}{2} \ve_{\a\b} \bar{\nabla}^{\bd i} \bar{W}_{\ad\bd} \mathbb{D} - \frac{\ri}{4} \ve_{\a\b} \bar{\nabla}^{\bd i} \bar{W}_{\ad\bd} Y + \ri \ve_{\a\b} \bar{\nabla}^\bd_j \bar{W}_{\ad\bd} J^{ij}
\eol & \quad
- \ri \ve_{\a\b} \bar{\nabla}_\bd^i \bar{W}_{\gd\ad} \bar{M}^{\bd \gd} - \frac{\ri}{4} \ve_{\a\b} \bar{\nabla}_\ad^i \bar{\nabla}^\bd_k \bar{W}_{\bd\gd} \bar{S}^{\gd k} + \frac{1}{2} \ve_{\a\b} \nabla^{\g \bd} \bar{W}_{\ad\bd} S^i_\g
\eol & \quad
+ \frac{\ri}{4} \ve_{\a\b} \bar{\nabla}_\ad^i \nabla^\g{}_\gd \bar{W}^{\gd \bd} K_{\g \bd}~.
\end{align}
\end{subequations}	
We also find that $W_{\a \b}$ obeys the Bianchi identity
\begin{align}
B = \nabla_{\a\b} W^{\a\b} &= \bar{\nabla}^{\ad\bd} \bar{W}_{\ad\bd}  = \bar{B} ~,
\end{align}
where $B$ is the $\N=2$ super-Bach tensor.
The equation of motion for $\cN=2$ conformal supergravity is the Bach-flatness 
condition
\bea
\nabla_{\a\b} W^{\a\b} &= \bar{\nabla}^{\ad\bd} \bar{W}_{\ad\bd}  = 0~.
\label{A.15}
\eea
We point out that the $\cN=1$ super-Bach tensor was introduced in \cite{BK88}, 
see also \cite{KMT,KP,BK}.


\begin{footnotesize}

\end{footnotesize}


\end{document}